\documentclass[draft]{agujournal2019}
\draftfalse
\journalname{Geophysical Research Letters}

\usepackage[hidelinks,final]{hyperref}

\makeatletter\def\NAT@sort{\z@}\makeatother

\begin{document}

\title{Radial Evolution of a CIR: Observations from a Nearly Radially Aligned Event Between Parker Solar Probe and STEREO-A}

\authors{R. C. Allen\affil{1}, G. C. Ho\affil{1}, G. M. Mason\affil{1}, G. Li\affil{2}, L. K. Jian\affil{3}, S. K. Vines\affil{1}, N. A. Schwadron\affil{4}, C. J. Joyce\affil{5}, S. D. Bale\affil{6,7,8,9}, J. W. Bonnell\affil{7}, A. W. Case\affil{10}, E. R. Christian\affil{3}, C. M. S. Cohen\affil{11}, M. I. Desai\affil{12}, R. Filwett\affil{13}, K. Goetz\affil{14}, P. R. Harvey\affil{7}, M. E. Hill\affil{1}, J. C. Kasper\affil{10,15}, K. E. Korreck\affil{10}, D. Lario\affil{3}, D. Larson\affil{7}, R. Livi\affil{7}, R. J. MacDowall\affil{3}, D. M. Malaspina\affil{16}, D. J. McComas\affil{5}, R. McNutt\affil{1}, D. G. Mitchell\affil{1}, K. W. Paulson\affil{10}, M. Pulupa\affil{7}, N. Raouafi\affil{1}, M. L. Stevens\affil{10}, P. L. Whittlesey\affil{7}, and M. Wiedenbeck\affil{17}}

\affiliation{1}{Johns Hopkins University, Applied Physics Lab, Laurel, MD 20723, USA.}
\affiliation{2}{Department of Space Science, University of Alabama in Huntsville, Huntsville, AL 35899, USA.}
\affiliation{3}{NASA/Goddard Space Flight Center, Greenbelt, MD 20771, USA.}
\affiliation{4}{University of New Hampshire, Space Science Center, Durham, NH, USA}
\affiliation{5}{Department of Astrophysical Science, Princeton University, Princeton, NH, USA}
\affiliation{6}{Physics Department, University of California, Berkeley, CA 94720, USA}
\affiliation{7}{Space Sciences Laboratory, University of California, Berkeley, CA 94720, USA}
\affiliation{8}{The Blackett Laboratory, Imperial College London, London, SW7 2AZ, UK}
\affiliation{9}{School of Physics and Astronomy, Queen Mary University of London, London E1 4NS, UK}
\affiliation{10}{Smithsonian Astrophysical Observatory, Cambridge, MA 02138, USA}
\affiliation{11}{California Institute of Technology, Pasadena, CA 91125, USA}
\affiliation{12}{Space Science and Engineering Division, Southwest Research Institute, San Antonio, TX, USA}
\affiliation{13}{Department of Physics and Astronomy, University of Iowa, Iowa City, Iowa 52242, USA}
\affiliation{14}{School of Physics and Astronomy, University of Minnesota, Minneapolis, MN 55455, USA}
\affiliation{15}{Climate and Space Sciences and Engineering, University of Michigan, Ann Arbor, MI 48109, USA}
\affiliation{16}{Laboratory for Atmospheric and Space Physics, University of Colorado, Boulder, CO 80303, USA}
\affiliation{17}{Jet Propulsion Laboratory, California Institute of Technology, Pasadena, CA, USA}

\correspondingauthor{Robert C. Allen}{Robert.Allen@jhuapl.edu}

\begin{keypoints}
\item A CIR was observed by PSP and STA when the spacecraft were near radially aligned.
\item The plasma measurements suggest only radial evolution has occurred between observations.
\item Differences in fast and slow solar wind magnetic topology lead to a wider longitudinal extent of the suprathermal ion enhancement at PSP.
\end{keypoints}

\begin{abstract}
The addition of Parker Solar Probe (PSP) to the Heliophysics System Observatory has allowed for the unprecedented ability to study Corotating Interaction Regions (CIRs) at multiple radial distances without significant temporal/longitudinal variations. On 2019 September 19, PSP observed a CIR at $\sim$0.5 au when it was nearly radially aligned with the Solar Terrestrial Relations Observatory-Ahead (STEREO-A) spacecraft at $\sim$1 au, allowing for an unambiguous assessment of the radial evolution of a single CIR. Bulk plasma and magnetic field signatures of the CIR evolve in a fashion characteristic to previous observations, however the suprathermal ions are enhanced over a larger longitudinal range at PSP than at STEREO-A, although at much lower intensities. The longitudinal spread appears to be largely a consequence of magnetic field line topology at CIRs between the compressed slow solar wind upstream and high-speed stream following the CIR, underscoring the importance of the large-scale topology of these structures.
\end{abstract}

\section{Introduction}

Corotating Interaction Regions (CIRs) form where corotating high speed streams, originating from coronal holes, overtake slower-speed solar wind (\citealp{Belcher1971}; for a recent review, see \citealp{Richardson2018}). As the solar wind flows radially outward, forward and reverse shocks can form along the CIR, often beyond 1 au (e.g., \citealp{Smith1976}; \citealp{Pizzo1978}; \citealp{Jian2006}; \citeyear{Jian2008}), which serve as a significant source of energetic particles in the interplanetary medium \citep{VanHollebeke1981,Tsurutani1982}. While shock acceleration has historically been viewed as a main acceleration mechanism at CIRs, some studies have indicated the importance of particle acceleration mechanisms within 1 au such as stochastic processes (e.g., \citealp{Richardson1985,Schwadron1996,Chotoo2000}) or processes in the unshocked compression region due to the velocity gradient across the CIR (e.g., \citealp{Giacalone2002,Ebert2012,Chen2015,Filwett2017}).

The Helios 1 and 2 missions, launched in 1974 and 1976, respectively, allowed the ability to probe CIRs in the inner heliosphere for the first time. To better understand how CIRs evolve with radial distance, \citet{Richter1986} performed a superposed epoch analysis using 16 CIRs between 0.3 and 0.4 au, and 31 CIRs between 0.9 and 1.0 au. They found that the density, temperature, pressure, and magnetic field increase at the CIR interface increased with increasing radial distance. The azimuthal variation in the solar wind velocity profile also steepened with increasing distance. Other studies of near radially-aligned CIR observations between the Helios spacecraft also found this velocity steepening and increased compression along the CIR interface (see \citealp{Balogh1999,Forsyth1999}). While these studies illustrate the evolution of the CIR structure with distance due to the increasing inclination of the Parker spiral with respect to the radial direction, other studies have found the opposite correlations. Statistical results by \citet{Schwenn1990}, for example, reported a decrease in the longitudinal speed gradients at the leading edge of CIRs from 0.3 to 0.5 au, attributing this to the sharp speed gradients at the boundaries of coronal holes, before becoming relatively unchanged from 0.5 to 1 au. Comparisons between Pioneer Venus Orbiter, and Advanced Composition Explorer (ACE)/Wind have also allowed for insight into the average variation of CIR characteristics between 0.72 au and 1 au \citep{Jian2008b}, finding little change in the velocity variation between these observations. However, very few events have been recorded by spacecraft at different radial distances but with the same longitude. As such, while changes in the average CIR profile by radial distance have been studied, though resulting in different trends at times, how individual CIRs evolve with radial distance is still largely unknown.

In addition to radial evolution, CIRs also evolve in time, longitude, and latitude. From one solar rotation to the next, the observed bulk plasma properties and the suprathermal particle intensities in CIRs change with time (e.g., \citealp{Mason2009}). Even when observed in close enough proximity, the suprathermal ion flux and spectra can be different between observations. Several studies have suggested this difference could be a result of successive observations being magnetically connected to a shock front at increasingly further distances, allowing for increasingly larger amounts of acceleration (e.g., \citealp{Barnes1976,Simnett1995,Zhao2015,Wijsen2019}). Beyond single CIR observations, solar cycle variations have also been observed in CIR occurrence and properties (\citealp{Jian2011}; \citeyear{Jian2019}) and in their suprathermal ion composition (\citealp{Mason2008}; \citeyear{Mason2012}; \citealp{Filwett2017}; \citealp{Allen2019}). As such, inferences from combining all CIR observations from many different points in time may not accurately represent any single constituent event. In the modern era of the Heliophysics System Observatory, Parker Solar Probe (PSP), in combination with observations at 1 au, allows for an unprecedented opportunity to disentangle the radial and temporal evolution of CIRs. During the first orbit of PSP, several studies investigated CIRs in the inner heliosphere \citep{McComas2019,Joyce2020,Desai2020,Cohen2020,Allen2020a}. \citet{Allen2020a} matched CIRs observed at PSP during its first orbit with observations at 1 au from the Solar Terrestrial Relations Observatory-Ahead (STEREO-A), ACE, and Wind missions. While this provided insight into possible differences in the energization of suprathermal ions in the inner heliosphere and their connectivity to shocks further out in the heliosphere, the conjunction geometry did not allow for the ability to differentiate between temporal and radial evolution.

In this study, we investigate a conjunction between PSP and STEREO-A, when the two spacecraft were nearly radially aligned. This allows for a comparison of the radial evolution of a CIR without the uncertainties arising from effects of significant temporal evolution. Section 2 describes the datasets used, the results of which are presented in Section 3. Discussion and summary of the conclusions of this work are given in Section 4.

\section{Missions and Datasets}

\subsection{Parker Solar Probe}

The PSP mission \citep{Fox2016} was launched into a heliocentric orbit around the Sun on 2018 August 12. This study uses 1-minute averages of bulk solar wind measurements from the Solar Probe Cup (SPC; \citealp{Case2020}), part of the Solar Wind Electrons Alpha and Protons (SWEAP) instrument suite \citep{Kasper2016}. Magnetic field measurements are provided by the FIELDS suite \citep{Bale2016}, and are averaged into a 1-min resolution product. This study computes proton temperature (T), specific entropy argument (S), and the combined proton plasma and magnetic pressure (P) following the methodology outlined in \citet{Allen2020a}. Suprathermal ion measurements from the Energetic Particle Instrument-Lo (EPI-Lo; \citealp{Hill2017}), part of the Integrated Science Investigation of the Sun (IS$\odot$IS; \citealp{McComas2016}), are averaged to a 30-minute resolution dataset.

\subsection{STEREO-A}

The STEREO mission \citep{Kaiser2008} is a set of two spacecraft that were launched on 2006 October 25, and sent to orbit the Sun in opposing directions when viewed in a Sun-Earth fixed frame. This study focuses on observations by STEREO-A (STA) and uses 1-minute averaged magnetic field observations from the magnetometer \citep{Acuna2008}, along with suprathermal particle measurements from the Solar Electron and Proton Telescope (SEPT; \citealp{MullerMellin2008}) instrument, both part of the In situ Measurements of Particles and CME Transients (IMPACT) investigation \citep{Luhmann2008}. SEPT is able to measure ions, but cannot differentiate ion species, so these ions are assumed to predominantly be H\textsuperscript{+}. Additionally, 1-minute bulk solar wind properties (i.e., velocity, temperature, and density) are provided by the Plasma and Suprathermal Ion Composition (PLASTIC) investigation \citep{Galvin2008}.

\subsection{Wind}

For comparison to STEREO-A, data from the Wind mission \citep{Acuna1995} are also presented. Wind was launched on 1994 November 1 and has been stationed at L1 since 2004. 1-minute resolution bulk solar wind plasma properties, from the Solar Wind Experiment (SWE; \citealp{Ogilvie1995}), and magnetic field measurements, from the Magnetic Field Instrument (MFI; \citealp{Lepping1995}), are used in this study.

\section{Results}

\subsection{Observations at 1 au}

A CIR was observed at 1 au by STA and Wind ($\sim$84$^{\circ}$ in longitude apart) during solar rotations prior to and after the observation of the CIR at PSP at 0.51 au on 2019 September 19. Figure 1 shows plasma, magnetic field and suprathermal particle observations at 1 au for this CIR. These observations are time-shifted such that the individual interfaces all line up with that of the third STA observation, denoted by the vertical dashed line in Figure 1. With each successive observation (comparing the dark blue and red traces in Figure 1a-h, in particular), the velocity increase becomes steeper, the density pileup becomes more pronounced, and the increases in temperature and entropy also steepen. Additionally, the suprathermal ion enhancement seen by STA (Figure 1i-k) becomes more intense. These variations between successive observations highlight the various ways in which long-lived CIRs can evolve in time.

\begin{figure}[tbp]
\centering
\includegraphics[width=\textwidth,height=0.8\textheight,keepaspectratio]{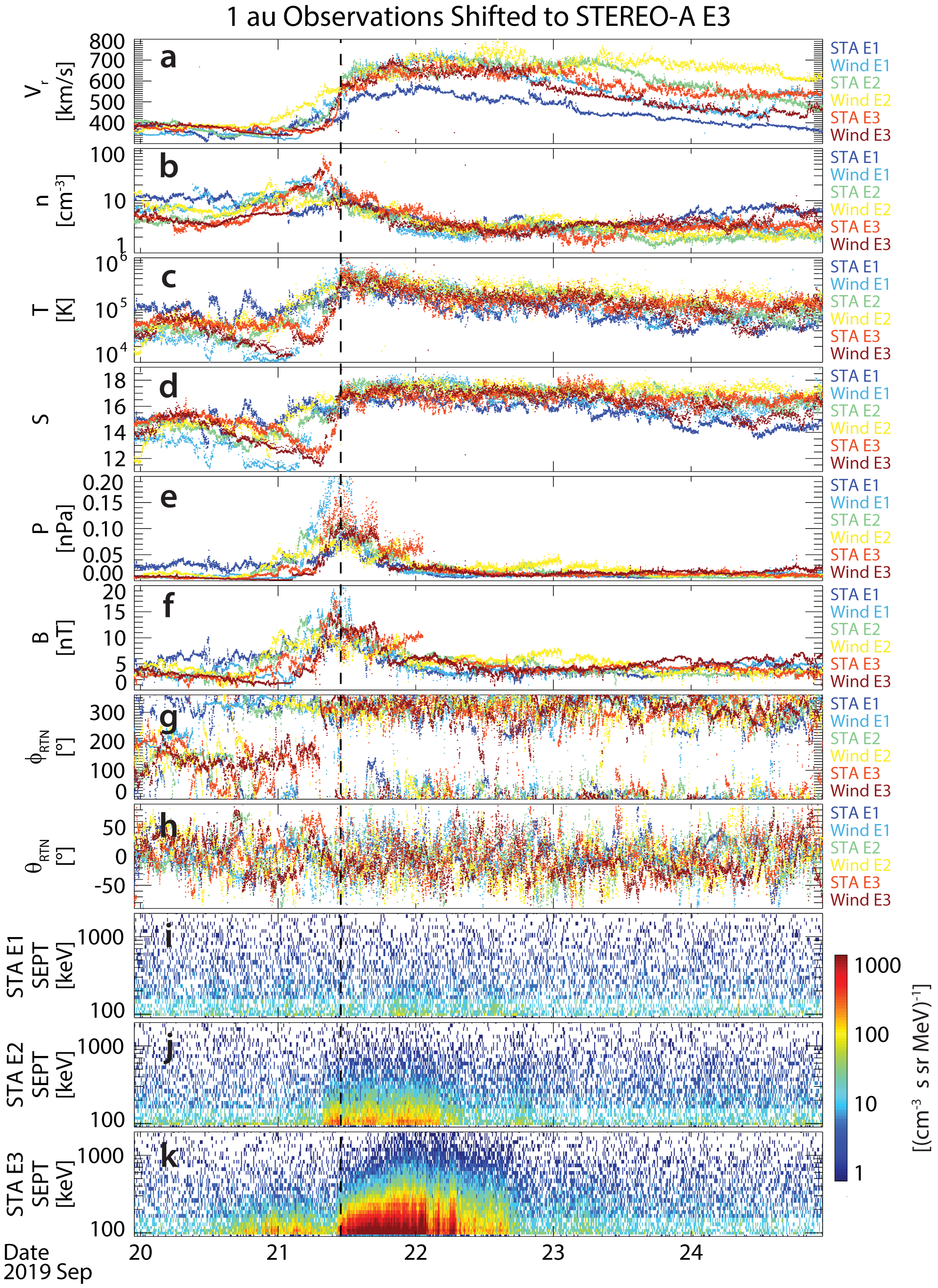}
\caption{Observations of a CIR at 1 au from STA and Wind time-shifted to match the stream interface of the STA event 3 (E3) (vertical dashed line). The CIR interface, identified as the time of peak pressure of the CIR (e.g., \citealp{Jian2006}), occurred at 2019-07-30/19:33:30 (labeled STA E1), 2019-08-05/08:01:30 (Wind E1), 2019-08-25/05:05:30 (STA E2), 2019-08-31/01:06:30 (Wind E2), 2019-09-21/10:55:30 (STA E3), and 2019-09-27/13:54:30 (Wind E3). From top to bottom, (a) radial velocity, (b) density, (c) temperature, (d) entropy, (e) combined proton and magnetic field pressure, (f) magnetic field magnitude, magnetic field (g) $\phi_{RTN}$ and (h) $\theta_{RTN}$ components in the RTN coordinate system, and (i-k) suprathermal ion intensity from SEPT on STA for the first, second, and third STA observations, respectively.}
\label{fig:1}
\end{figure}

\subsection{Radial evolution}

During the third STA observation of this CIR (STA E3 in Figure 1), STA and PSP were nearly radially aligned (within 4.2$^{\circ}$ longitude and 0.26$^{\circ}$ latitude when the CIR passed over PSP on 2019 September 19\textsuperscript{th}), with PSP at $\sim$0.5 au ($\sim$80$^{\circ}$ in longitude from L1) and STA at 0.98 au. To compare the solar wind and IMF observations at different radial distances (\emph{r}), approximate scaling laws (see \citealp{Kivelson1995}) were applied to the STA dataset: $n \propto r^{-2}$, $T \propto r^{-4/3}$, $B_{r} \propto r^{-2}$, and $B_{t} \propto r^{-1}$, where \emph{B}\textsubscript{r} and \emph{B}\textsubscript{t} denote the radial and tangential components of the interplanetary magnetic field (IMF) in radial-tangential-normal (RTN) coordinates. These scalings are applied to remove the underlying systematic radial variations, such as from volumetric expansion, to highlight the variations caused by the interaction between the slow and high-speed streams. Figure 2 illustrates the PSP observations (in black) and the scaled STA observations (in blue) that are also shifted in time by -1.77 days to align the CIR interfaces (as determined by eye). This shift is consistent with corotation after taking into account the different radial distances. The velocity, density, temperature, entropy, pressure, and magnetic field observations at PSP and the radially-scaled STA E3 observations are in remarkable agreement, suggesting very little temporal variation between the observations. Focusing on the transition from slow to fast solar wind, the density pileup and pressure enhancement are more pronounced at STA than at PSP, although comparisons of the velocity, temperature, entropy, and magnetic field strength are difficult to fully assess due to data gaps during the interval at PSP (to conserve power, the instruments are powered off during scheduled high-rate Ka-band downlink opportunities). While both PSP and STA are within 0.26$^{\circ}$ of each other in heliographic latitude, STA likely observed the CIR closer to the heliospheric current sheet (HCS) than PSP (as apparent in the abrupt change in $\phi_{RTN}$ observed just before the CIR; Figure 2h). Additionally, a suprathermal ion flux enhancement associated with the CIR was observed at both PSP (Figure 2i) and STA (Figure 2j). The small enhancement at STA prior to the HCS crossing is likely unrelated to the CIR (see, \citealp{Smith1978}). Notably, the suprathermal ion enhancement at PSP is observed for a longer duration, corresponding to a larger longitudinal extent, than at STA.

\begin{figure}[tbp]
\centering
\includegraphics[width=\textwidth,height=0.8\textheight,keepaspectratio]{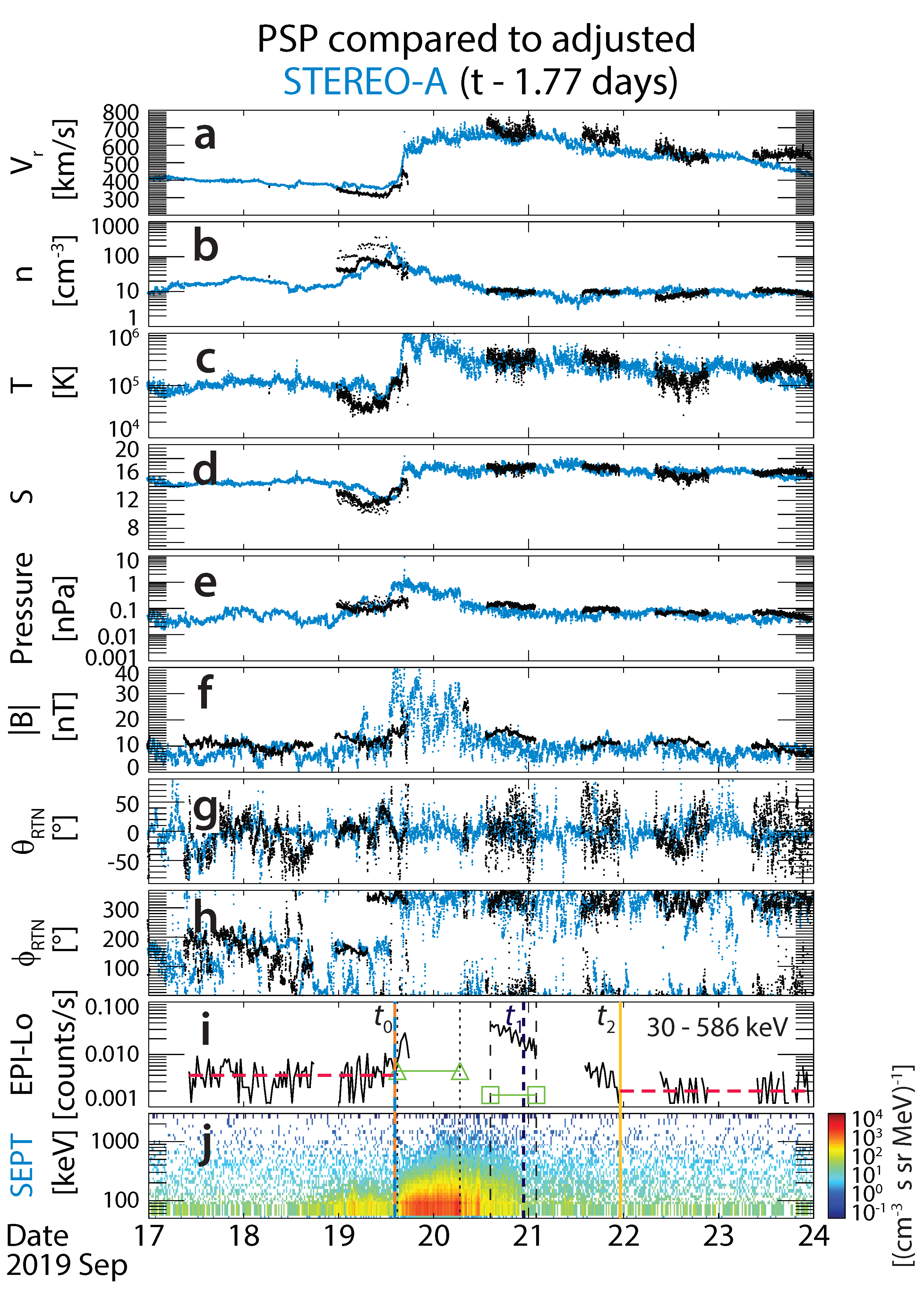}
\caption{CIR observation at PSP (black) and STA (blue) for STA E3. The STA E3 observations are shifted in time by 1.77 days and are scaled according to the relations provided in the text. From top to bottom, (a) radial velocity, (b) density, (c) temperature, (d) entropy, (e) combined proton and magnetic field pressure, (f) magnetic field magnitude, (g) $\theta_{RTN}$, (h) $\phi_{RTN}$, (i) PSP EPI-Lo ``ToF only'' mode count rate, and (j) STA SEPT flux spectrograms are shown. See text for description of markings on panels i and j, denoting intervals in Figure 3 and Figure 4.}
\label{fig:2}
\end{figure}

To further investigate the radial change of the suprathermal ion population, intensities over two time intervals, 2019-09-19/14:40 -- 2019-09-20/06:40 UT (defined to be within the main pressure enhancement of the CIR and denoted by the dotted vertical lines and green triangles) and 2019-09-20/14:00 -- 2019-09-21/02:00 UT (encompassing the time period in the high speed stream with both PSP and STA suprathermal ion observations and denoted by the vertical dashed lines and green squares), were averaged to construct ion spectra. Due to the data gaps at PSP, EPI-Lo observations are only shown for the second timeframe. Figure 3 displays these spectra with STA SEPT observations (green), PSP H\textsuperscript{+} (red), and PSP \textsuperscript{4}He\textsuperscript{n+} (blue). Power law fits were applied for energies $<$1000 keV/nuc for STA and $<$600 keV/nuc for PSP. The enhancement in suprathermal ion flux for PSP H\textsuperscript{+} above 600 keV is related to known cross-talk background in the instrument (see \citealp{Hill2020}), and is ignored in our analysis.

\begin{figure}[tbp]
\centering
\includegraphics[width=0.7\textwidth]{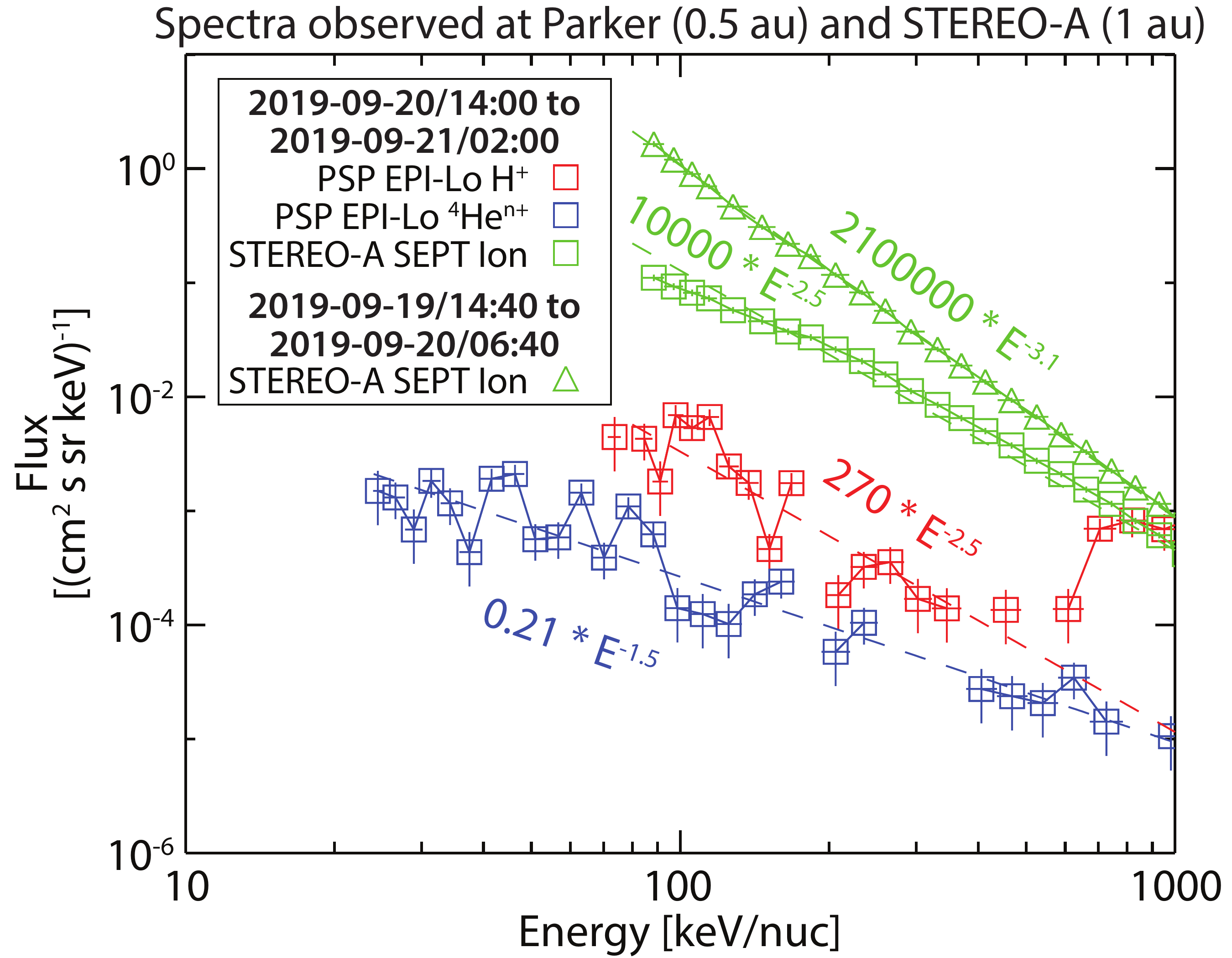}
\caption{Flux spectra for suprathermal \textsuperscript{4}He\textsuperscript{n+} (blue) and H\textsuperscript{+} (red) observed at PSP, and total ions (green) observed at STA for a sub-interval inside the CIR structure (triangles) and within the trailing high-speed stream (squares) where both STA and PSP have measurements available.}
\label{fig:3}
\end{figure}

Comparing the STA ion power law within the CIR (triangles) to those later in the high speed stream (squares), the power law hardens slightly through the event (-3.1 to -2.5, respectively). While the power law indices are nearly the same between PSP and the later STA interval (-2.5), the intensity of the suprathermal ion enhancement at PSP is significantly lower by a factor of $\sim$40 (i.e., red versus green squares in Figure 3).

\subsection{Longitudinal Spread of Energetic Particles}

To further investigate the longitudinal spread of the suprathermal ion enhancement at PSP versus that at STA, we define start and stop times of the enhancements for both observations. Red horizontal dashed lines on Figure 2i mark the average EPI-Lo count rates before and after the suprathermal ion enhancement at PSP. The time that the EPI-Lo count rate exceeds the averaged pre-event background count rate is considered the start time (t\textsubscript{0}=2019-09-19/14:00 UT, shown by the orange vertical dashed line in Figure 2i-j). Using a similar method, the time-shifted STA SEPT flux was found to become enhanced at the same time (in the shifted timeframe). The time the STA count rate reached the post-event average rate was defined as the STA enhancement stop time (t\textsubscript{1}=2019-09-20/22:20 UT, shown by the thick vertical dashed line in Figure 2i-j). This method also identified the PSP enhancement stop time (t\textsubscript{2}, shown as the solid yellow vertical line in Figure 2i-j) at 2019-09-21/23:30 UT. Comparing the times of the suprathermal ion enhancements at STA to the bulk plasma properties during this interval, the enhancements begin near the CIR interface. Additionally, there is a decrease in the suprathermal ion intensity near the trailing edge of the CIR at STA, however the intensity remains elevated into the high-speed stream. Due to data gaps in the PSP observations, we cannot reliably compare the energetic particle enhancements at PSP to the bulk plasma structure of the CIR in the same fashion.

To further examine the role of the underlying IMF structure in the longitudinal spread of the energetic particles, Parker spiral IMF lines were computed using the solar wind velocity observed by PSP at the beginning and end of the suprathermal ion enhancements (t\textsubscript{0} and t\textsubscript{2} computed above). The Parker spiral IMF line corresponding to the end of the suprathermal ion enhancement at PSP was then corotated backward to correspond to its location at the start of the ion enhancement (t\textsubscript{0}). For both the Parker spiral calculations and corotation distance, a fixed corotation speed of 14.7$^{\circ}$/day was used, consistent with the equatorial corotation speed of the Sun. Using a wider range of corotation speeds that have been previously observed for CIRs (13.4$^{\circ}$/day -- 14.7$^{\circ}$/day, cf. \citealp{Allen2020b}) yielded similar results (not shown). The two Parker spiral IMF lines are shown in Figure 4, with the ``compression field line'' denoting the Parker spiral based on observations at the start of the ion enhancement at PSP (t\textsubscript{0}), and the ``rarefaction field line'' denoting the corotated Parker spiral in the fast solar wind when the suprathermal ion flux returned to the background level (t\textsubscript{2}).

\begin{figure}[tbp]
\centering
\includegraphics[width=0.85\textwidth]{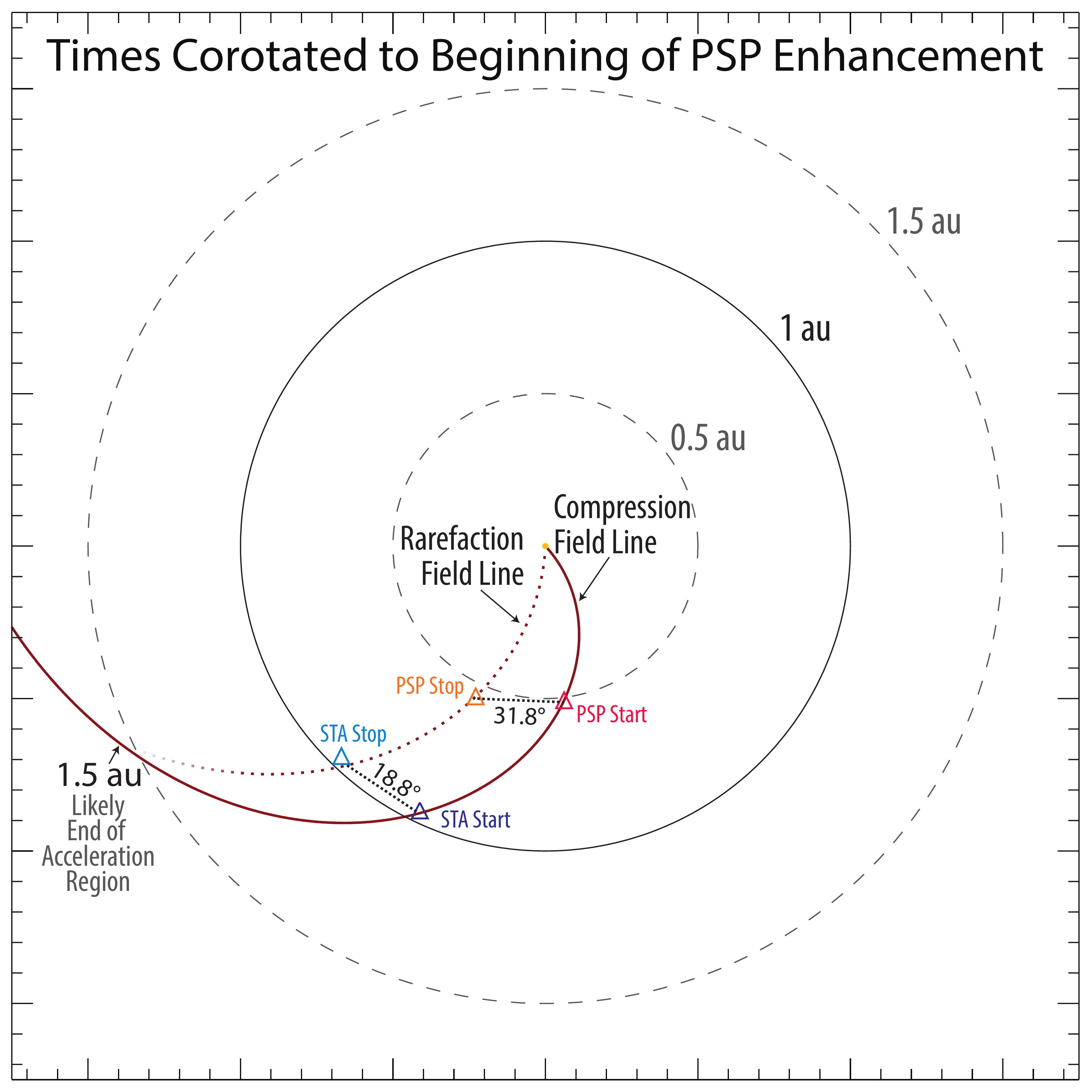}
\caption{Parker spiral field lines calculated from PSP observations at the start and stop times of the suprathermal ion enhancements are shown by the solid and dotted maroon lines, respectively. The start and stop location of the PSP and STA suprathermal ion enhancements (triangles) are corotated to the beginning of the PSP enhancement.}
\label{fig:4}
\end{figure}

Taking the start/stop times of the suprathermal ion enhancements at both PSP and STA, we compute the time difference between those times and t\textsubscript{0}. Using this time difference and a fixed corotation speed (14.7$^{\circ}$/day), we corotate the locations of PSP and STA observations to the reference time t\textsubscript{0}, resulting in the labeled triangles in Figure 4. These locations are seen to largely agree with the Parker spiral field lines. Due to the rarefaction field line being more radial than the compression field line, arising from the different solar wind speeds at those times, the different curvatures lead to variable longitudinal spreads. For this event, the longitudinal spread at the radial distance of PSP was 31.8$^{\circ}$, while the longitudinal spread at STA was only 18.8$^{\circ}$. While it is known that the interaction between slow and fast streams can cause the magnetic topology to deviate from a Parker spiral configuration, this coarse approximation over smaller distances (0.5 au in this case) is found to be appropriate in aligning these observations. Although the zeroth order treatment here aligns the timing of the observations quite well, small variations are observed which could be an effect of higher-order corrections to the Parker spiral and/or differences in instrument sensitivities. The compression and rarefaction field lines approach each other at a distance of $\sim$1.5 au, suggesting the CIR-associated acceleration processes are occurring within this distance, and so allowing the suprathermal particles access to the flux tube constrained by these field lines.

\section{Discussion and Summary}

This study explores a unique and fortuitous CIR event when PSP and STA were nearly radially aligned. Additionally, we examined the variations of this CIR over several solar rotations as seen by STA and Wind. At 1 au, the CIR structure persisted for three solar rotations and shows a clear temporal evolution from one observation to the next (Figure 1). The third STA observation of the CIR occurred when STA and PSP were nearly radially aligned. Shifting the STA E3 measurements earlier in time by 1.77 days temporally aligned the observations between PSP and STA. Scaling the STA plasma and fields measurements by theoretical radial dependencies aligned the observations remarkably, suggesting little temporal evolution between observations at PSP and the third STA observation of this CIR (Figure 2).

Examining the suprathermal ion energy spectra associated with the CIR observed at PSP, the \textsuperscript{4}He\textsuperscript{n+} spectral slope is harder than for H\textsuperscript{+} (-1.5 to -2.5, respectively). This is often observed for CIRs at and beyond 1 au, and is thought to be an effect of preferential acceleration of pick up ions at CIRs, possibly indicating acceleration within 1 au (e.g., \citealp{Gloeckler1994,Schwadron1996,Gloeckler1998,Chotoo2000}). Future studies probing how the spectral slopes for helium ions as compared to protons vary with radial distance can give further insight into species-dependent transport and acceleration processes. Looking at the full CIR interval at 1 au, the STA ion spectral slope evolves throughout the event (noting the green triangles versus the green squares in Figure 3). This evolution in the spectral slopes is likely associated with the spectra being taken inside versus outside of the CIR structure (e.g., \citealp{Barnes1976}). When observations exist for both spacecraft towards the latter half of the suprathermal ion enhancement, the spectral indices of ions at STA and H\textsuperscript{+} at PSP are approximately the same ($\sim$2.5; green and red squares for STA and PSP, respectively, in Figure 3).

Under the assumption that the main source of particle acceleration is at heliocentric distances beyond both spacecraft, transport processes undergone by the particles would typically be invoked to explain lower particle intensities with decreasing heliocentric distances. While the suprathermal H\textsuperscript{+} intensities are $\sim$40 times lower at PSP than STA during the time period following the CIR interface when both spacecraft have data, the spectral profiles at both spacecraft have the same slope. Because these processes are also predicted to manifest themselves as an increasingly hardened spectra at lower energies with distance from the source region (e.g., \citealp{Fisk1980}), but no such hardening is observed in this event, this points to far weaker modulation of energetic particles than expected, as also indicated by other studies (e.g., \citealp{Mason1997}; \citealp{Mason1999}; \citealp{Mason2008}; \citeyear{Mason2012}; \citealp{Desai2020}; \citealp{Schwadron2020}).

While the suprathermal ion enhancement at PSP is observed over a longer time period and longitudinal range than that seen for STA, this can be explained due to the geometry of the magnetic field lines (Figure 4). For this event, the suprathermal ion enhancement at both PSP and STA begin in the compressed slow solar wind and end in the rarefaction region high speed stream (Figure 2). As a result of the difference in solar wind speed, the Parker spiral field line in the high-speed stream is more radial than that of the compressed slow solar wind. This difference in Parker spiral geometry results in a radial dependence to the longitudinal extent of the flux tube. For this event, this corresponds to a longitudinal extent of 18.8$^{\circ}$ and 31.8$^{\circ}$ at the radial distance of STA and PSP, respectively, consistent with the observations. Under the assumption that the particles were accelerated at larger heliospheric distances, the width of the suprathermal ion enhancements suggests acceleration occurred within $\sim$1.5 au. Since STA did not observe a shock at 1 au, but only a developing reverse shock, it is not clear if a shock became fully formed at slightly further heliospheric distances, or if instead compressive acceleration is resulting in the observed suprathermal particles through stochastic processes and/or mechanisms occurring in the unshocked compression region due to the velocity gradient across the CIR, similar to diffusive shock acceleration at a quasi-parallel shock (e.g., \citealp{Richardson1985,Chotoo2000,Giacalone2002,Ebert2012,Chen2015,Filwett2017,Joyce2020}). Due to the close proximity to the acceleration region estimated for this event (within 1.5 au), the Parker spiral approximation for the magnetic field topology well describes the general structuring of this CIR event to zeroth-order. However, determining the topology of CIR events in which acceleration occurs at larger heliospheric distances would require more detailed mapping due to sub-Parker spiral field lines (e.g., \citealp{Murphy2002}; \citealp{Schwadron2002}; \citealp{Schwadron2005}, \citealp{Schwadron2020}) and the interaction between the slow and fast solar wind along the CIR interface. Further investigation of other events between PSP and other observatories can be used to better understand the transport of suprathermal particles into the inner heliosphere. Additionally, future conjunctions between PSP, STA, Wind, and Solar Orbiter will allow for continued investigation into the variations of CIRs with radial distance, longitude, and latitude.

\acknowledgments
This work was supported by NASA's Parker Solar Probe Mission, contract NNN06AA01C. Parker Solar Probe data can be accessed from \url{https://sppgway.jhuapl.edu/}. The STEREO SEPT data are available at \url{http://www2.physik.uni-kiel.de/stereo/data/sept/}, and STEREO magnetic field and plasma data can be found at the STEREO Science Center: \url{https://stereo-ssc.nascom.nasa.gov}. Wind data are available at \url{https://cdaweb.gsfc.nasa.gov/}. This event is taken from the catalog by \citet{Allen2020c}.

\bibliography{psp_sta_cir}

\begin{thebibliography}{}
\expandafter\ifx\csname natexlab\endcsname\relax\def\natexlab#1{#1}\fi
\providecommand{\url}[1]{\href{#1}{#1}}
\providecommand{\dodoi}[1]{doi:~\href{http://doi.org/#1}{\nolinkurl{#1}}}
\providecommand{\doeprint}[1]{\href{http://ascl.net/#1}{\nolinkurl{http://ascl.net/#1}}}
\providecommand{\doarXiv}[1]{\href{https://arxiv.org/abs/#1}{\nolinkurl{https://arxiv.org/abs/#1}}}

\bibitem[{Acu{\~n}a {et~al.}(2008)Acu{\~n}a, Curtis, Scheifele, Russell,
  Schroeder, Szabo, \& Luhmann}]{Acuna2008}
Acu{\~n}a, M.~H., Curtis, D., Scheifele, J.~L., {et~al.} 2008, Space Science
  Reviews, 136, 203, \dodoi{10.1007/s11214-007-9259-2}

\bibitem[{Acu{\~n}a {et~al.}(1995)Acu{\~n}a, Ogilvie, Baker, Curtis, Fairfield,
  \& Mish}]{Acuna1995}
Acu{\~n}a, M.~H., Ogilvie, K.~W., Baker, D.~N., {et~al.} 1995, Space Science
  Reviews, 71, 5, \dodoi{10.1007/BF00751323}

\bibitem[{Allen {et~al.}(2020b)Allen, Ho, Jian, Mason, Vines, \&
  Lario}]{Allen2020b}
Allen, R.~C., Ho, G.~C., Jian, L.~K., {et~al.} 2020b, Space Weather, 18,
  \dodoi{10.1029/2019SW002437}

\bibitem[{Allen {et~al.}(2019)Allen, Ho, \& Mason}]{Allen2019}
Allen, R.~C., Ho, G.~C., \& Mason, G.~M. 2019, The Astrophysical Journal
  Letters, 883, L10, \dodoi{10.3847/2041-8213/ab3f2f}

\bibitem[{Allen {et~al.}(2020a)Allen, Lario, Odstrcil, Ho, Jian, Cohen, Badman,
  Jones, Arge, Mays, Mason, Bale, Bonnell, Case, Christian, de~Wit, Goetz,
  Harvey, Henney, Hill, Kasper, Korreck, Larson, Livi, MacDowall, Malaspina,
  McComas, McNutt, Mitchell, Pulupa, Raouafi, Schwadron, Stevens, Whittlesey,
  \& Wiedenbeck}]{Allen2020a}
Allen, R.~C., Lario, D., Odstrcil, D., {et~al.} 2020a, The Astrophysical
  Journal Supplement Series, 246, 36, \dodoi{10.3847/1538-4365/ab578f}

\bibitem[{Allen {et~al.}(2020c)Allen, Ho, Jian, Vines, Bale, Case, Hill, Joyce,
  Kasper, Korreck, Malaspina, McComas, McNutt, M{\"o}stl, Odstrcil, Raouafi,
  Schwadron, \& Stevens}]{Allen2020c}
Allen, R.~C., Ho, G.~C., Jian, L.~K., {et~al.} 2020c, Astronomy \&
  Astrophysics, 650, A25, \dodoi{10.1051/0004-6361/202039833}

\bibitem[{Bale {et~al.}(2016)Bale, Goetz, Harvey, Turin, Bonnell, Dudok~de Wit,
  Ergun, MacDowall, Pulupa, Andre, Bolton, Bougeret, Bowen, Burgess, Cattell,
  Chandran, Chaston, Chen, Choi, Connerney, Cranmer, Diaz-Aguado, Donakowski,
  Drake, Farrell, Fergeau, Fermin, Fischer, Fox, Glaser, Goldstein, Gordon,
  Hanson, Harris, Hayes, Hinze, Hollweg, Horbury, Howard, Hoxie, Jannet,
  Karlsson, Kasper, Kellogg, Kien, Klimchuk, Krasnoselskikh, Krucker, Lynch,
  Maksimovic, Malaspina, Marker, Martin, Martinez-Oliveros, McCauley, McComas,
  McDonald, Meyer-Vernet, Moncuquet, Monson, Mozer, Murphy, Odom, Oliverson,
  Olson, Parker, Pankow, Phan, Quataert, Quinn, Ruplin, Salem, Seitz, Sheppard,
  Siy, Stevens, Summers, Szabo, Timofeeva, Vaivads, Velli, Yehle, Werthimer, \&
  Wygant}]{Bale2016}
Bale, S.~D., Goetz, K., Harvey, P.~R., {et~al.} 2016, Space Science Reviews,
  204, 49, \dodoi{10.1007/s11214-016-0244-5}

\bibitem[{Balogh {et~al.}(1999)Balogh, Bothmer, Crooker, Forsyth, Gloeckler,
  Hewish, Hilchenbach, Kallenbach, Klecker, Linker, Lucek, Mann, Marsch,
  Posner, Richardson, Schmidt, Scholer, Wang, Wimmer-Schweingruber, Aellig,
  Bochsler, Hefti, \& Miki{\'c}}]{Balogh1999}
Balogh, A., Bothmer, V., Crooker, N.~U., {et~al.} 1999, Space Science Reviews,
  89, 141, \dodoi{10.1023/A:1005245306874}

\bibitem[{Barnes \& Simpson(1976)}]{Barnes1976}
Barnes, C.~W., \& Simpson, J.~A. 1976, \apjl, 210, L91, \dodoi{10.1086/182311}

\bibitem[{Belcher \& Davis(1971)}]{Belcher1971}
Belcher, J.~W., \& Davis, L. 1971, Journal of Geophysical Research, 76, 3534,
  \dodoi{10.1029/JA076i016p03534}

\bibitem[{Case {et~al.}(2020)Case, Kasper, Stevens, Korreck, Paulson, Daigneau,
  Caldwell, Freeman, Henry, Klingensmith, Bookbinder, Robinson, Berg, Tiu,
  Wright, Reinhart, Curtis, Ludlam, Larson, Whittlesey, Livi, Klein, \&
  Martinovi{\'c}}]{Case2020}
Case, A.~W., Kasper, J.~C., Stevens, M.~L., {et~al.} 2020, The Astrophysical
  Journal Supplement Series, 246, 43, \dodoi{10.3847/1538-4365/ab5a7b}

\bibitem[{Chen {et~al.}(2015)Chen, Schwadron, M{\"o}bius, \& Gorby}]{Chen2015}
Chen, J.~H., Schwadron, N.~A., M{\"o}bius, E., \& Gorby, M. 2015, Journal of
  Geophysical Research: Space Physics, 120, 9269, \dodoi{10.1002/2014JA020939}

\bibitem[{Chotoo {et~al.}(2000)Chotoo, Schwadron, Mason, Zurbuchen, Gloeckler,
  Posner, Fisk, Galvin, Hamilton, \& Collier}]{Chotoo2000}
Chotoo, K., Schwadron, N.~A., Mason, G.~M., {et~al.} 2000, Journal of
  Geophysical Research: Space Physics, 105, 23107, \dodoi{10.1029/1998JA000015}

\bibitem[{Cohen {et~al.}(2020)Cohen, Christian, Cummings, Davis, Desai,
  Giacalone, Hill, Joyce, Labrador, Leske, Matthaeus, McComas, McNutt, Mewaldt,
  Mitchell, Rankin, Roelof, Schwadron, Stone, Szalay, Wiedenbeck, Allen, Ho,
  Jian, Lario, Odstrcil, Bale, Badman, Pulupa, MacDowall, Kasper, Case,
  Korreck, Larson, Livi, Stevens, \& Whittlesey}]{Cohen2020}
Cohen, C. M.~S., Christian, E.~R., Cummings, A.~C., {et~al.} 2020, The
  Astrophysical Journal Supplement Series, 246, 20,
  \dodoi{10.3847/1538-4365/ab4c38}

\bibitem[{Desai {et~al.}(2020)Desai, Mitchell, Szalay, Roelof, Giacalone, Hill,
  McComas, Christian, Schwadron, McNutt~Jr., Wiedenbeck, Joyce, Cohen, Ebert,
  Dayeh, Allen, Davis, Krimigis, Leske, Matthaeus, Malandraki, Mewaldt,
  Labrador, Stone, Bale, Pulupa, MacDowall, \& Kasper}]{Desai2020}
Desai, M.~I., Mitchell, D.~G., Szalay, J.~R., {et~al.} 2020, The Astrophysical
  Journal Supplement Series, 246, 56, \dodoi{10.3847/1538-4365/ab65ef}

\bibitem[{Ebert {et~al.}(2012)Ebert, Dayeh, Desai, \& Mason}]{Ebert2012}
Ebert, R.~W., Dayeh, M.~A., Desai, M.~I., \& Mason, G.~M. 2012, The
  Astrophysical Journal, 749, 73, \dodoi{10.1088/0004-637X/749/1/73}

\bibitem[{Filwett {et~al.}(2017)Filwett, Desai, Dayeh, \&
  Broiles}]{Filwett2017}
Filwett, R.~J., Desai, M.~I., Dayeh, M.~A., \& Broiles, T.~W. 2017, The
  Astrophysical Journal, 838, 23, \dodoi{10.3847/1538-4357/aa5ca9}

\bibitem[{Fisk \& Lee(1980)}]{Fisk1980}
Fisk, L.~A., \& Lee, M.~A. 1980, The Astrophysical Journal, 237, 620,
  \dodoi{10.1086/157907}

\bibitem[{Forsyth \& Marsch(1999)}]{Forsyth1999}
Forsyth, R.~J., \& Marsch, E. 1999, Space Science Reviews, 89, 7,
  \dodoi{10.1023/A:1005235626013}

\bibitem[{Fox {et~al.}(2016)Fox, Velli, Bale, Decker, Driesman, Howard, Kasper,
  Kinnison, Kusterer, Lario, Lockwood, McComas, Raouafi, \& Szabo}]{Fox2016}
Fox, N.~J., Velli, M.~C., Bale, S.~D., {et~al.} 2016, Space Science Reviews,
  204, 7, \dodoi{10.1007/s11214-015-0211-6}

\bibitem[{Galvin {et~al.}(2008)Galvin, Kistler, Popecki, Farrugia, Simunac,
  Ellis, M{\"o}bius, Lee, Boehm, Carroll, Crawshaw, Conti, Demaine, Ellis,
  Gaidos, Googins, Granoff, Gustafson, Heirtzler, King, Knauss, Levasseur,
  Longworth, Singer, Turco, Vachon, Vosbury, Widholm, Blush, Karrer, Bochsler,
  Daoudi, Etter, Fischer, Jost, Opitz, Sigrist, Wurz, Klecker, Ertl,
  Seidenschwang, Wimmer-Schweingruber, Koeten, Thompson, \&
  Steinfeld}]{Galvin2008}
Galvin, A.~B., Kistler, L.~M., Popecki, M.~A., {et~al.} 2008, Space Science
  Reviews, 136, 437, \dodoi{10.1007/s11214-007-9296-x}

\bibitem[{Giacalone {et~al.}(2002)Giacalone, Jokipii, \& Kota}]{Giacalone2002}
Giacalone, J., Jokipii, J.~R., \& Kota, J. 2002, The Astrophysical Journal,
  573, 845, \dodoi{10.1086/340660}

\bibitem[{Gloeckler \& Geiss(1998)}]{Gloeckler1998}
Gloeckler, G., \& Geiss, J. 1998, Space Science Reviews, 86, 127,
  \dodoi{10.1023/A:1005019628054}

\bibitem[{Gloeckler {et~al.}(1994)Gloeckler, Geiss, Roelof, Fisk, Ipavich,
  Ogilvie, Lanzerotti, von Steiger, \& Wilken}]{Gloeckler1994}
Gloeckler, G., Geiss, J., Roelof, E.~C., {et~al.} 1994, Journal of Geophysical
  Research: Space Physics, 99, 17637, \dodoi{10.1029/94JA01509}

\bibitem[{Hill {et~al.}(2017)Hill, Mitchell, Andrews, Cooper, Gurnee, Hayes,
  Layman, McNutt, Nelson, Parker, Schlemm, Stokes, Begley, Boyle, Burgum, Do,
  Dupont, Gold, Haggerty, Hoffer, Hutcheson, Jaskulek, Krimigis, Liang, London,
  Noble, Roelof, Seifert, Strohbehn, Vandegriff, \& Westlake}]{Hill2017}
Hill, M.~E., Mitchell, D.~G., Andrews, G.~B., {et~al.} 2017, Journal of
  Geophysical Research: Space Physics, 122, 1513, \dodoi{10.1002/2016JA022614}

\bibitem[{Hill {et~al.}(2020)Hill, Mitchell, Allen, de~Nolfo, Vourlidas, Brown,
  Jones, McComas, McNutt, Mitchell, Szalay, Wallace, Arge, Christian, Cohen,
  Crew, Desai, Giacalone, Henney, Joyce, Krimigis, Leske, Mewaldt, Nelson,
  Roelof, Schwadron, \& Wiedenbeck}]{Hill2020}
Hill, M.~E., Mitchell, D.~G., Allen, R.~C., {et~al.} 2020, The Astrophysical
  Journal Supplement Series, 246, 65, \dodoi{10.3847/1538-4365/ab643d}

\bibitem[{Jian {et~al.}(2006)Jian, Russell, Luhmann, \& Skoug}]{Jian2006}
Jian, L., Russell, C.~T., Luhmann, J.~G., \& Skoug, R.~M. 2006, Solar Physics,
  239, 337, \dodoi{10.1007/s11207-006-0132-3}

\bibitem[{Jian {et~al.}(2019)Jian, Luhmann, Russell, \& Galvin}]{Jian2019}
Jian, L.~K., Luhmann, J.~G., Russell, C.~T., \& Galvin, A.~B. 2019, Solar
  Physics, 294, \dodoi{10.1007/s11207-019-1416-8}

\bibitem[{Jian {et~al.}(2011)Jian, Russell, \& Luhmann}]{Jian2011}
Jian, L.~K., Russell, C.~T., \& Luhmann, J.~G. 2011, Solar Physics, 274, 321,
  \dodoi{10.1007/s11207-011-9737-2}

\bibitem[{Jian {et~al.}(2008{\natexlab{a}})Jian, Russell, Luhmann, Skoug, \&
  Steinberg}]{Jian2008}
Jian, L.~K., Russell, C.~T., Luhmann, J.~G., Skoug, R.~M., \& Steinberg, J.~T.
  2008{\natexlab{a}}, Solar Physics, 250, 375,
  \dodoi{10.1007/s11207-008-9204-x}

\bibitem[{Jian {et~al.}(2008{\natexlab{b}})Jian, Russell, Luhmann, Skoug, \&
  Steinberg}]{Jian2008b}
---. 2008{\natexlab{b}}, Solar Physics, 249, 85,
  \dodoi{10.1007/s11207-008-9161-4}

\bibitem[{Joyce {et~al.}(2020)Joyce, McComas, Christian, Schwadron, Wiedenbeck,
  McNutt, Cohen, Leske, Mewaldt, Stone, Labrador, Davis, Cummings, Mitchell,
  Hill, Roelof, Szalay, Rankin, Desai, Giacalone, \& Matthaeus}]{Joyce2020}
Joyce, C.~J., McComas, D.~J., Christian, E.~R., {et~al.} 2020, The
  Astrophysical Journal Supplement Series, 246, 41,
  \dodoi{10.3847/1538-4365/ab5948}

\bibitem[{Kaiser {et~al.}(2008)Kaiser, Kucera, Davila, St.~Cyr, Guhathakurta,
  \& Christian}]{Kaiser2008}
Kaiser, M.~L., Kucera, T.~A., Davila, J.~M., {et~al.} 2008, Space Science
  Reviews, 136, 5, \dodoi{10.1007/s11214-007-9277-0}

\bibitem[{Kasper {et~al.}(2016)Kasper, Abiad, Austin, Balat-Pichelin, Bale,
  Belcher, Berg, Bergner, Berthomier, Bookbinder, Brodu, Caldwell, Case,
  Chandran, Cheimets, Cirtain, Cranmer, Curtis, Daigneau, Dalton, Dasgupta,
  DeTomaso, Diaz-Aguado, Djordjevic, Donaskowski, Effinger, Florinski, Fox,
  Freeman, Gallagher, Gary, Gauron, Gates, Goldstein, Golub, Gordon, Gurnee,
  Guth, Halekas, Hatch, Heerikuisen, Ho, Hu, Johnson, Jordan, Korreck, Larson,
  Lazarus, Li, Livi, Ludlam, Maksimovic, McFadden, Marchant, Maruca, McComas,
  Messina, Mercer, Park, Peddie, Pogorelov, Reinhart, Richardson, Robinson,
  Rosen, Skoug, Slagle, Steinberg, Stevens, Szabo, Taylor, Tiu, Turin, Velli,
  Webb, Whittlesey, Wright, Wu, \& Zank}]{Kasper2016}
Kasper, J.~C., Abiad, R., Austin, G., {et~al.} 2016, Space Science Reviews,
  204, 131, \dodoi{10.1007/s11214-015-0206-3}

\bibitem[{Kivelson \& Russell(1995)}]{Kivelson1995}
Kivelson, M.~G., \& Russell, C.~T. 1995, Introduction to Space Physics (New
  York, NY: Cambridge University Press)

\bibitem[{Lepping {et~al.}(1995)Lepping, Acu{\~n}a, Burlaga, Farrell, Slavin,
  Schatten, Mariani, Ness, Neubauer, Whang, Byrnes, Kennon, Panetta, Scheifele,
  \& Worley}]{Lepping1995}
Lepping, R.~P., Acu{\~n}a, M.~H., Burlaga, L.~F., {et~al.} 1995, Space Science
  Reviews, 71, 207, \dodoi{10.1007/BF00751330}

\bibitem[{Luhmann {et~al.}(2008)Luhmann, Curtis, Schroeder, McCauley, Lin,
  Larson, Bale, Sauvaud, Aoustin, Mewaldt, Cummings, Stone, Davis, Cook,
  Kecman, Wiedenbeck, von Rosenvinge, Acuna, Reichenthal, Shuman, Wortman,
  Reames, Mueller-Mellin, Kunow, Mason, Walpole, Korth, Sanderson, Russell, \&
  Gosling}]{Luhmann2008}
Luhmann, J.~G., Curtis, D.~W., Schroeder, P., {et~al.} 2008, Space Science
  Reviews, 136, 117, \dodoi{10.1007/s11214-007-9170-x}

\bibitem[{Mason {et~al.}(2012)Mason, Desai, \& Li}]{Mason2012}
Mason, G.~M., Desai, M.~I., \& Li, G. 2012, The Astrophysical Journal, 748,
  L31, \dodoi{10.1088/2041-8205/748/2/L31}

\bibitem[{Mason {et~al.}(2009)Mason, Desai, Mall, Korth, Bucik, von Rosenvinge,
  \& Simunac}]{Mason2009}
Mason, G.~M., Desai, M.~I., Mall, U., {et~al.} 2009, Solar Physics, 256, 393,
  \dodoi{10.1007/s11207-009-9367-0}

\bibitem[{Mason {et~al.}(1997)Mason, Mazur, Dwyer, Reames, \& von
  Rosenvinge}]{Mason1997}
Mason, G.~M., Mazur, J.~E., Dwyer, J.~R., Reames, D.~V., \& von Rosenvinge,
  T.~T. 1997, The Astrophysical Journal, 486, L149, \dodoi{10.1086/310845}

\bibitem[{Mason \& Sanderson(1999)}]{Mason1999}
Mason, G.~M., \& Sanderson, T.~R. 1999, Space Science Reviews, 89, 77,
  \dodoi{10.1023/A:1005216516443}

\bibitem[{Mason {et~al.}(2008)Mason, Leske, Desai, Cohen, Dwyer, Mazur,
  Mewaldt, Gold, \& Krimigis}]{Mason2008}
Mason, G.~M., Leske, R.~A., Desai, M.~I., {et~al.} 2008, The Astrophysical
  Journal, 678, 1458, \dodoi{10.1086/533524}

\bibitem[{McComas {et~al.}(2016)McComas, Alexander, Angold, Bale, Beebe,
  Birdwell, Boyle, Burgum, Burnham, Christian, Cook, Cooper, Cummings, Davis,
  Desai, Dickinson, Dirks, Do, Fox, Giacalone, Gold, Gurnee, Hayes, Hill,
  Kasper, Kecman, Klemic, Krimigis, Labrador, Layman, Leske, Livi, Matthaeus,
  McNutt, Mewaldt, Mitchell, Nelson, Parker, Rankin, Roelof, Schwadron,
  Seifert, Shuman, Stokes, Stone, Vandegriff, Velli, von Rosenvinge, Weidner,
  Wiedenbeck, \& Wilson}]{McComas2016}
McComas, D.~J., Alexander, N., Angold, N., {et~al.} 2016, Space Science
  Reviews, 204, 187, \dodoi{10.1007/s11214-014-0059-1}

\bibitem[{McComas {et~al.}(2019)McComas, Christian, Cohen, Cummings, Davis,
  Desai, Giacalone, Hill, Joyce, Krimigis, Labrador, Leske, Malandraki,
  Matthaeus, McNutt, Mewaldt, Mitchell, Posner, Rankin, Roelof, Schwadron,
  Stone, Szalay, Wiedenbeck, Bale, Kasper, Case, Korreck, MacDowall, Pulupa,
  Stevens, \& Rouillard}]{McComas2019}
McComas, D.~J., Christian, E.~R., Cohen, C. M.~S., {et~al.} 2019, Nature, 576,
  223, \dodoi{10.1038/s41586-019-1811-1}

\bibitem[{M{\"u}ller-Mellin {et~al.}(2008)M{\"u}ller-Mellin, B{\"o}ttcher,
  Falenski, Rode, Duvet, Sanderson, Butler, Johlander, \&
  Smit}]{MullerMellin2008}
M{\"u}ller-Mellin, R., B{\"o}ttcher, S., Falenski, J., {et~al.} 2008, Space
  Science Reviews, 136, 363, \dodoi{10.1007/s11214-007-9204-4}

\bibitem[{Murphy {et~al.}(2002)Murphy, Smith, \& Schwadron}]{Murphy2002}
Murphy, N., Smith, E.~J., \& Schwadron, N.~A. 2002, Geophysical Research
  Letters, 29, \dodoi{10.1029/2002GL015164}

\bibitem[{Ogilvie {et~al.}(1995)Ogilvie, Chornay, Fritzenreiter, Hunsaker,
  Keller, Lobell, Miller, Scudder, Sittler, Torbert, Bodet, Needell, Lazarus,
  Steinberg, Tappan, Mavretic, \& Gergin}]{Ogilvie1995}
Ogilvie, K.~W., Chornay, D.~J., Fritzenreiter, R.~J., {et~al.} 1995, Space
  Science Reviews, 71, 55, \dodoi{10.1007/BF00751326}

\bibitem[{Pizzo(1978)}]{Pizzo1978}
Pizzo, V. 1978, Journal of Geophysical Research: Space Physics, 83, 5563,
  \dodoi{10.1029/JA083iA12p05563}

\bibitem[{Richardson(1985)}]{Richardson1985}
Richardson, I.~G. 1985, Planetary and Space Science, 33, 557,
  \dodoi{10.1016/0032-0633(85)90097-2}

\bibitem[{Richardson(2018)}]{Richardson2018}
---. 2018, Living Reviews in Solar Physics, 15,
  \dodoi{10.1007/s41116-017-0011-z}

\bibitem[{Richter \& Luttrell(1986)}]{Richter1986}
Richter, A.~K., \& Luttrell, A.~H. 1986, Journal of Geophysical Research: Space
  Physics, 91, 5873, \dodoi{10.1029/JA091iA05p05873}

\bibitem[{Schwadron(2002)}]{Schwadron2002}
Schwadron, N.~A. 2002, Geophysical Research Letters, 29,
  \dodoi{10.1029/2002GL015028}

\bibitem[{Schwadron {et~al.}(1996)Schwadron, Fisk, \&
  Gloeckler}]{Schwadron1996}
Schwadron, N.~A., Fisk, L.~A., \& Gloeckler, G. 1996, Geophysical Research
  Letters, 23, 2871, \dodoi{10.1029/96GL02833}

\bibitem[{Schwadron \& McComas(2005)}]{Schwadron2005}
Schwadron, N.~A., \& McComas, D.~J. 2005, Geophysical Research Letters, 32,
  \dodoi{10.1029/2004GL021579}

\bibitem[{Schwadron {et~al.}(2020)Schwadron, Joyce, Aly, Cohen, Desai, McComas,
  Niehof, M{\"o}bius, Lee, Bower, Bale, Case, Christian, Davis, de~Wet, Goetz,
  Giacalone, Hill, Allen, Kasper, Korreck, Leske, Malandraki, Matthaeus,
  McNutt, Mewaldt, Mitchell, Pulupa, Rankin, Roelof, Stone, Szalay, \&
  Wiedenbeck}]{Schwadron2020}
Schwadron, N.~A., Joyce, C.~J., Aly, A., {et~al.} 2020, Astronomy \&
  Astrophysics, 650, A24, \dodoi{10.1051/0004-6361/202039352}

\bibitem[{Schwenn(1990)}]{Schwenn1990}
Schwenn, R. 1990, in Physics of the Inner Heliosphere I (Springer Berlin
  Heidelberg), 99--181, \dodoi{10.1007/978-3-642-75361-9_3}

\bibitem[{Simnett \& Roelof(1995)}]{Simnett1995}
Simnett, G.~M., \& Roelof, E.~C. 1995, \ssr, 72, 303,
  \dodoi{10.1007/BF00768795}

\bibitem[{Smith {et~al.}(1978)Smith, Tsurutani, \& Rosenberg}]{Smith1978}
Smith, E.~J., Tsurutani, B.~T., \& Rosenberg, R.~L. 1978, Journal of
  Geophysical Research: Space Physics, 83, 717, \dodoi{10.1029/JA083iA02p00717}

\bibitem[{Smith \& Wolfe(1976)}]{Smith1976}
Smith, E.~J., \& Wolfe, J.~H. 1976, Geophysical Research Letters, 3, 137,
  \dodoi{10.1029/GL003i003p00137}

\bibitem[{Tsurutani {et~al.}(1982)Tsurutani, Smith, Pyle, \&
  Simpson}]{Tsurutani1982}
Tsurutani, B.~T., Smith, E.~J., Pyle, K.~R., \& Simpson, J.~A. 1982, Journal of
  Geophysical Research: Space Physics, 87, 7389,
  \dodoi{10.1029/JA087iA09p07389}

\bibitem[{van Hollebeke {et~al.}(1981)van Hollebeke, McDonald, Trainor, \& von
  Rosenvinge}]{VanHollebeke1981}
van Hollebeke, M. A.~I., McDonald, F.~B., Trainor, J.~H., \& von Rosenvinge,
  T.~T. 1981, in Solar Wind 4, Proceedings of the Conference held August
  18--September 1, 1978 in Burghausen, FDR, ed. H.~Rosenbauer, MPAE-W-100-81-31
  (Garching, FDR: Max-Planck-Institut f{\"u}r Aeronomie), 497

\bibitem[{Wijsen {et~al.}(2019)Wijsen, Aran, Pomoell, \& Poedts}]{Wijsen2019}
Wijsen, N., Aran, A., Pomoell, J., \& Poedts, S. 2019, \aap, 624, A47,
  \dodoi{10.1051/0004-6361/201935139}

\bibitem[{Zhao {et~al.}(2015)Zhao, Li, Ebert, Dayeh, Desai, Mason, Wu, \&
  Chen}]{Zhao2015}
Zhao, L., Li, G., Ebert, R.~W., {et~al.} 2015, Journal of Geophysical Research:
  Space Physics, 121, 77, \dodoi{10.1002/2015JA021762}

\end{thebibliography}

\end{document}